\documentclass[twocolumn, preprintnumbers, aps]{revtex4}
\usepackage{epsfig,amssymb,amsmath}

\newcommand{\bi}{\bibitem}
\newcommand{\be}{\begin{eqnarray}}
\newcommand{\ee}{\end{eqnarray}}
\newcommand{\rar}{\rightarrow}

\begin{document}

\title{A revision of the Generalized Uncertainty Principle}

\author{Cosimo Bambi}

\affiliation{Department of Physics and Astronomy, 
Wayne State University, Detroit, MI 48201, USA}

\date{\today}

\preprint{WSU-HEP-0806}

\begin{abstract}
The Generalized Uncertainty Principle arises from the 
Heisenberg Uncertainty Principle when gravity is taken 
into account, so the leading order correction to the 
standard formula is expected to be proportional to the 
gravitational constant $G_N = L_{Pl}^2$. On the other 
hand, the emerging picture suggests a set of departures
from the standard theory which demand a revision of all
the arguments used to deduce heuristically the new rule.
In particular, one can now argue that the leading order 
correction to the Heisenberg Uncertainty Principle is 
proportional to the first power of the Planck length 
$L_{Pl}$. If so, the departures from ordinary quantum 
mechanics would be much less suppressed than what is 
commonly thought.
\end{abstract}

\maketitle

\section{Introduction}

The search for a common description of particle physics and 
gravity and for a quantum theory of the gravitational sector 
is certainly one of the most outstanding and longstanding 
problems in physics. Despite important discoveries, at present 
a reliable theoretical framework is lacking and even the very 
meaning of quantum spacetime is not clear. However, from 
heuristic considerations, quite general arguments and simple 
analogies, we expect a set of model independent features which 
should occur in any consistent theory of quantum gravity. 
Among other things, it is common belief that the Heisenberg 
Uncertainty Principle (HUP) $\Delta x \, \Delta p \gtrsim 1$ 
has to be replaced by the so--called Generalized Uncertainty 
Principle (GUP)~\cite{string, bh, gr}
\be\label{gup}
\Delta x \, \Delta p \gtrsim 1 
+ \alpha \, L_{Pl}^2 \, (\Delta p)^2 \, ,
\ee
where $\alpha$ is a positive dimensionless coefficient (which
in general may depend on $x$ and $p$) and $L_{Pl} \sim 10^{-33}$~cm
is the Planck length. Since $\alpha > 0$, eq.~(\ref{gup})
implies the existence of a minimum observable length, very
probably close to the Planck one. On the other hand, the GUP is 
deduced from heuristic arguments which use standard concepts of 
different frameworks, such as quantum mechanics from one hand and 
general relativity from the other hand, so the final result can 
be at most ``reasonable'', but cannot be ``fully reliable''. In 
particular, one should see eq.~(\ref{gup}) as a first approximation 
to a more complex inequality, whose exact form is probably 
impossible to derive from general grounds and outside a well 
defined theory.

An important feature of eq.~(\ref{gup}) is that there is no term
linear in the Planck length. This is exactly what one can find 
in the literature, since no derivation provides it: the reason is 
basically that the modification of the HUP arises when gravitational 
interactions are taken into account, so the leading order 
correction is expected to to be proportional to the gravitational 
constant $G_N = L_{Pl}^2$. On the other hand, the GUP implies the 
existence of a minimum length of order $L_{Pl}$ and the latter is
promoted to fundamental parameter in the new picture. Revising the
derivation of the GUP, it is now natural to expand the GUP in
powers of $L_{Pl}$, rather than of $G_N$, and to include the term 
linear in $L_{Pl}$ into eq.~(\ref{gup}). We can arrive at the 
same conclusion by noticing that the GUP requires a modification 
of the standard de Broglie relation (indeed wavelength smaller 
than the minimum length cannot make sense) and this is always 
neglect when the GUP is derived, because at a first approach 
one uses standard physics only. On the other hand, a modified de 
Broglie relation could, in turn, generate a term proportional to 
$L_{Pl}$ in the GUP. This is basically the result of the present 
work. If so, the departures from standard quantum mechanics would 
be much less suppressed than what is commonly believed and most 
consequences of the GUP discussed in the literature would have 
to be reconsidered.

The content of the paper is as follows. In secs.~\ref{sec-hup}
and \ref{sec-gup}, I review the basic ingredients which are
used respectively in the derivation of the HUP and of the GUP. 
In sec.~\ref{sec-rev}, I show that the leading order correction 
to the HUP could be proportional to the first power of the 
Planck length. In secs.~\ref{sec-atom} and \ref{sec-ho}, 
I briefly discuss the implications of the proposal of the present 
work with two simple examples. In sec.~\ref{sec-concl}, there are 
summary and conclusions.

\section{Heisenberg Uncertainty Principle \label{sec-hup}}

\begin{figure}[t]
\par
\begin{center}
\includegraphics[width=6.5cm,angle=0]{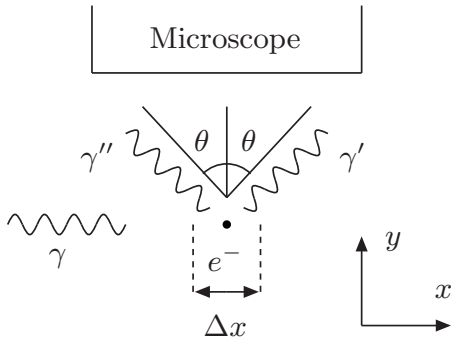}
\end{center}
\par
\vspace{-7mm} 
\caption{Heisenberg's microscope: experimental setup.}
\label{fig}
\end{figure}

The Heisenberg's microscope is a gedanken experiment whose
purpose is to measure position and momentum of a particle, say
an electron, by using some probe, for example a photon. The 
setup of the experiment is sketched in fig.~\ref{fig}. Since 
the photon has a wavelength $\lambda$, we are able to resolve 
at best length scales of order $\lambda$ itself. More precisely,
wave optics theory requires that the projected electron 
position uncertainty is~\footnote{For the sake of simplicity, 
throughout the paper I focus the attention on one space dimension 
only. In the case of two or more space dimensions, the exact 
picture becomes more complicated and model dependent, but the 
basic features are essentially the same. I always use 
$\hbar = c = 1$ units.}
\be\label{hup-x}
\Delta x \gtrsim \frac{\lambda}{2 \theta} \, , 
\ee
where $\theta \ll 1$ is the angle defined in fig.~\ref{fig}. Of
course, in order to be detected by the microscope, the photon
must be scattered within the cone of angle $2\theta$. The two
extreme cases are reported in fig.~\ref{fig}, where the scattered
photon is indicated respectively with $\gamma'$ and $\gamma''$.
In the first case, the photon hits the right edge of the lens
of the microscope: the momentum in the $x$ direction of the
electron after the collision is $p'_x$, while the one of the
photon is $q' \sin\theta$, where $q'$ is the photon momentum.
In the second case, the photon hits the left edge of the lens 
of the microscope and we have respectively $p''_x$ for the
electron and $-q'' \sin\theta$ for the photon. Since the
total momentum of the system is conserved, the two limiting
cases have equal final momentum in the $x$ direction
\be\label{hup-0}
p'_x + q' \sin\theta = p''_x - q'' \sin\theta \, .
\ee 
For $\theta \ll 1$, $\sin\theta \approx \theta$ and $q' \approx q''$.
On the other hand, since we cannot distinguish experimentally
the two cases, the electron momentum uncertainty in the $x$
direction is
\be\label{hup-p}
\Delta p \gtrsim p''_x - p'_x \sim 2 q' \theta \, .
\ee
Lastly, we use the de Broglie relation to connect the momentum 
$q'$ of the photon with its wavelength $\lambda$
\be
q' = 2 \pi / \lambda
\ee
and we multiply eq.~(\ref{hup-x}) by eq.~(\ref{hup-p}), finding
the well known result $\Delta x \, \Delta p \gtrsim 1$. 
Let me remark the crucial assumptions leading to the HUP: wave
optics theory, momentum conservation and de Broglie relation.

\section{Generalized Uncertainty Principle \label{sec-gup}}

When one takes gravitational interactions into account, the
ordinary HUP must be revised, since the classical concept of 
spacetime breaks down as we approach distances close to the 
Planck length $L_{Pl}$. The result is the GUP in eq.~(\ref{gup}), 
which can be deduced heuristically from several different 
frameworks (string theories~\cite{string}, black hole physics~\cite{bh} 
and even more general considerations on the gravitational 
interactions~\cite{gr}), so it is common belief that it 
must hold in any consistent picture of quantum gravity. 
Consequences of the GUP can be found e.g. in ref.~\cite{gup-ph}.

A simple derivation of eq.~(\ref{gup}) can be obtained even
from dimensional grounds. In addition to the uncertainty due
to the wave--particle duality, there must be one caused by the
gravitational interaction between the electron and the photon, 
which increases as the photon energy increases. This is an 
inevitable effect that cannot be shielded and that is independent 
of the nature of the probe, as follows from the Equivalence 
Principle. Hence, the leading order of the extra uncertainty 
is at least proportional to the gravitational coupling constant 
$G_N = L_{Pl}^2$. Then, for dimensional reasons, one finds 
eq.~(\ref{gup}).

Another simple and intuitive derivation is the following. From
the ordinary HUP, we know that the electron position accuracy
could be improved using more and more energetic photons. However, 
it is also clear that this approach has a lower bound, because if 
we use too much energetic photons, a black hole is formed in the 
collision between the photon and the electron. This statement
can also be rephrased by saying that it is impossible to localize
an amount of energy with a space resolution better than the
radius of the black hole with the same mass. So, the position 
uncertainty cannot decrease below the corresponding black hole 
radius and the minimum space resolution is
\be\label{max}
(\Delta x)_{min} \sim \max \left[ \, 1/(\Delta p) \, , \, 
G_N \, \Delta p \, \right] \, .
\ee
Combining linearly the two uncertainties in eq.~(\ref{max}), we 
find the GUP of eq.~(\ref{gup}).

Even if it is often ignored, the GUP should be seen only as a
first approximation to a more complicated formula. Indeed, it
is deduced using standard physics, while it is well known 
that in the emerging picture several rules of our current
theoretical frameworks break down.

Before proceeding, let us notice two important points:
\begin{enumerate}
\item The basic ingredients which are used to find the
leading order correction to the HUP are the Equivalence 
Principle, according to which all the forms of energy 
couple to gravity with the same strength, and ordinary 
special relativity kinematics, in particular the standard 
dispersion relation for massless particles, i.e. $P = E$, 
and the standard energy--momentum conservation. In models 
where the minimum length is independent of the reference 
frame, at least the last two assumptions could be wrong, 
see e.g. ref.~\cite{dsr}. However, if we use standard 
expressions, we introduce only errors of order higher than 
$L_{Pl}^2$.
\item Since the modification of the HUP arises from the 
gravitational interaction between the photon and the 
electron, the leading order correction is expected to be 
at least proportional to $G_N$ or, in other words, there 
is no term linear in the Planck length. This is what one
finds in all the presently available literature on the GUP.
\end{enumerate}

\section{Revision of the GUP \label{sec-rev}}

Since the GUP implies the existence of a minimum length, the
wavelength of every particle has now a natural lower bound and 
a modification of the standard de Broglie relation seems inevitable,
see ref.~\cite{gup-mdr} and note~\footnote{In principle, 
there may be the possibility that the de Broglie relation 
holds up to a maximum momentum $p_{max}$ and minimum wavelength 
$\lambda_{min}$ (not smaller than the minimum length of the 
theory) and that the momentum of a single particle cannot exceed 
$p_{max}$. However, even if there exists this maximum momentum, as 
in the frameworks of ref.~\cite{dsr}, such an explanation cannot work 
for non--elementary objects.}. In particular, we can expect that 
the wavelength of a particle can be written as $\lambda = f(p)$, 
where $f(p)$ is some model dependent function of the momentum $p$ 
such that, at low energies ($p \ll L_{Pl}^{-1}$), we recover the 
standard de Broglie relation
\be
f(p) \approx \frac{2 \pi}{p} \, ,
\ee
while at high energies (beyond the Planck scale) $f(p)$ goes 
asymptotically to a minimum wavelength, quite naturally equal 
or somewhat larger than the minimum length of the theory,
\be
f(p) \rar \lambda_{min} \, .
\ee
So, for $p \ll L_{Pl}^{-1}$, we can write
\be\label{db-exp}
\lambda = \frac{2 \pi}{p} \, 
\left( 1 + a \, L_{Pl} \, p + b \, L_{Pl}^2 \, p^2 + ... \right) \, ,
\ee
where $a$, $b$,... are dimensionless model dependent parameters.

If we now repeat the argument leading to the HUP with the
new de Broglie relation $\lambda=f(p)$, we find
\be
\Delta x \, \Delta p &\gtrsim& q \, f(q) \sim \nonumber\\
&\sim& 1 + a \, L_{Pl} \, \Delta p 
+ b \, L_{Pl}^2 \, (\Delta p)^2 + ... 
\ee
and, in general, we can expect the term linear in the Planck 
length. Then, when we take gravity into account
\be\label{gup-l}
\Delta x \, \Delta p \gtrsim 1 
+ a \, L_{Pl} \, \Delta p 
+ \alpha' \, L_{Pl}^2 \, (\Delta p)^2 + ... \, .
\ee
In the simplest case $\alpha' = b + \alpha$, where
$b$ comes from the modified de Broglie relation, while $\alpha$
is due to the gravitational interaction between the electron 
and the photon. The ellipsis on the right hand side of 
eq.~(\ref{gup-l}) means that surely we are neglecting higher 
order corrections in $L_{Pl}$. Even if in principle the 
dimensionless parameter $a$ could be positive or
negative, we can reasonably guess that $f(p)$ is a monotonic
function and hence $a \ge 0$. As for the sign of higher order 
coefficients in eq.~(\ref{db-exp}), we do not have any 
indication, but anyway the expression must be consistent with 
the existence of a minimum length.

For instance, a modified de Broglie relation which leads to
$a \neq 0$ and seems consistent with general requirements is  
\be\label{db2}
\lambda & = & 
\frac{2 \pi L_{Pl}}{1 - \exp \left( - \, L_{Pl} \, p \right)}
= \nonumber\\
&=& \frac{2 \pi}{p} \left( 1 + 
\frac{1}{2} \, L_{Pl} \, p + ... \right) \, .
\ee
A common choice which can be found in the 
literature is instead
\be\label{db1}
\lambda & = & 
\frac{2 \pi L_{Pl}}{\tanh \left( L_{Pl} \, p \right)}
= \nonumber\\
&=& \frac{2 \pi}{p} \left( 1 + 
\frac{1}{3} \, L_{Pl}^2 \, p^2 + ... \right) \, .
\ee
Eq.~(\ref{db1}) is derived for example in ref.~\cite{kmm},
where the authors assume, among other things, the following
canonical commutation relation
\be\label{ccr}
[x,p] = i \left(1 + \alpha L_{Pl}^2 \, p^2 \right) \, .
\ee
Even if it is not the unique possibility, eq.~(\ref{ccr}) 
leads to the GUP in eq.~(\ref{gup}). However, if we relax 
the requirement that the new commutation relation has
to reproduce exactly the GUP in eq.~(\ref{gup}), because 
we believe that the latter is only an approximated expression,
both in the canonical commutation relation and in the de Broglie 
relation the leading order corrections to their standard 
expressions can be proportional to the first power of
the Planck length. Unfortunately, at least at present, there 
are no model independent arguments which can suggest a 
particular high energy generalization of the relation between 
momentum and wavelength and hence we have a large number of 
reasonable candidates.

The derivation of the modified de Broglie relation from the
assumed canonical commutation relation is non--trivial and
depends on several assumptions, in particular on the
fate of Lorentz invariance at short distances (since a minimum
length implies the deformation or violation of the latter) and
how it is realized in the model. However, if we knew the
``true'' commutation relation $[x,p]$, we would not need any
gedanken experiment to find the new uncertainty principle,
but we could use the general formula
\be\label{[ab]}
(\Delta A)^2 \, (\Delta B)^2 \ge 
\frac{1}{4} \left| \langle [A,B] \rangle \right|^2 \, ,
\ee
which holds for any observable $A$ and $B$ (in our case $A,B = x,p$)
and basically relies only on the definition of uncertainty
and expectation value of observables. From eq.~(\ref{[ab]})
we clearly see that $a \neq 0$ demands just that the leading 
order correction of $[x,p]$ is linear in $L_{Pl}$. If it is not,
then $a = 0$.

The outcome of the Heisenberg's experiment may be also affected
by a deviation from standard conservation rules. Indeed, even 
if not strictly compulsory, in some theoretical frameworks with 
a minimum length the usual energy--momentum conservation is 
``deformed''. An example can be found in the third paper 
of ref.~\cite{dsr}, where the authors consider a 2--body 
collision $1+2 \rar 3+4$ and find
\be
\frac{p_1}{1 - L_{m}E_1} + \frac{p_2}{1 - L_{m}E_2} 
= \frac{p_3}{1- L_{m}E_3} + \frac{p_4}{1 - L_{m}E_4}
\, , \nonumber\\
\ee 
where $L_{m}$ is the minimum and universal length scale of the
theory, which in our case is $L_{Pl}$ or of order $L_{Pl}$. 
Here the counterpart of eq.~(\ref{hup-0}) is
\be
p'_x (1 + L_m E' + ...) + q' \sin\theta (1 + L_m q' + ...) = 
\nonumber\\ \qquad
p''_x (1 + L_m E'' + ...) - q'' \sin\theta (1 + L_m q'' + ...) 
\, , \nonumber\\
\ee
where $E'$ and $E''$ are the electron energies after the 
collision with the photon in the two limiting cases. 
Eq.~(\ref{hup-p}) has now to be replaced with
\be
\Delta p &\gtrsim& p''_x - p'_x \sim 
\nonumber\\ &\sim& 2 q' \theta 
+ L_m (2 q'^2 \theta + p'_x E' - p''_x E'') + ... \, .
\ee
and we get eq.~(\ref{gup-l}) with $a L_{Pl} \sim L_m$
if there are no other corrections linear in $L_{Pl}$.

Here I would like to stress that when one deduces heuristically 
for the first time the GUP, the term proportional to $L_{Pl}$ 
cannot appear, because the Planck length has no intrinsic meaning 
in Newtonian gravity or in general relativity. On the other hand, 
in the emerging picture $L_{Pl}$ is promoted to fundamental 
parameter of the theory, which controls the minimum space 
resolution, and revising self--consistently the GUP, we cannot 
exclude the possibility that $a\neq 0$ in eq.~(\ref{gup-l}).
More precisely, it is clear from eq.~(\ref{max}) that the GUP
is obtained when we combine quantum mechanics, which provides 
a good description of physical phenomena at energies much 
smaller than the Planck scale, with standard gravity, which
works for macroscopic objects, whose masses are many orders of
magnitude larger than the Planck one. On the other hand, we
are unable to describe the microscopic physics near the Planck
scale and for this reason we do not know strong arguments which
require to write eq.~(\ref{gup-l}) with $a \neq 0$. One could
try to find indications from string theory, but even the
latter is not fully under control near the Planck (string) scale.
However there may be arguments which point to this direction: indeed, 
the usual description of black holes is expected to break down
below a minimum mass $M_{min} \sim M_s / g_s^2$, where $M_s$
is the string scale (usually somewhat smaller than the Planck 
mass) and $g_s$ is the string coupling, often assumed smaller 
than one. There may be an energy range around the Planck/string scale 
where physics is intrinsically stringy and before the interaction 
between the electron and the photon is disturbed by black hole 
production (see the second argument in sec.~\ref{sec-rev}),
the formation of ``string balls'' could dominate the uncertainty
of the electron position, see e.g.~\cite{s-ball} and references 
therein. These string balls would be objects whose size at the 
moment of their formation is at the level of the string length 
$L_s = 1/M_s$, independently of their mass (by contrast, black hole
radius increases linearly with the mass). If so, there
would be an energy range where the spatial resolution is limited
by the creation of such extended objects and we could be tempted
to replace eq.~(\ref{max}) with
\be\label{max-rev}
(\Delta x)_{min} \sim \max \left[ \, 1/(\Delta p) \, , 
\, L_s \, , G_N \, \Delta p \, \right] \, .
\ee
Then, combining linearly the three uncertainty in eq.~(\ref{max-rev}),
we would get eq.~(\ref{gup-l}) with $L_s \approx a L_{Pl}$ and
$\alpha' = 1$ ($L_s$ is usually only a little larger than $L_{Pl}$, 
say $L_s \sim L_{Pl}/g_s$).

\section{Hydrogen atom \label{sec-atom}}

It is well known that the hydrogen atom (and every atom in 
general) should quickly decay classically, but that the HUP 
makes it stable. Indeed, the energy of the electron orbiting 
around the proton is
\be\label{hydrogen}
E \sim \frac{p^2}{2 \, m_e} - \frac{e^2}{r} \, .
\ee
and classically it has no lower bound: for $r \rar 0$, 
$E \rar - \infty$. On the other hand, from the HUP follows
$p \gtrsim \Delta p \gtrsim 1/\Delta r \gtrsim 1/r$ and
therefore
\be
E \gtrsim \frac{1}{2 \, m_e \, r^2} 
- \frac{e^2}{r} \ge - \frac{m_e \, e^4}{2} 
= - E_0 \sim - 10 \; {\rm eV} \, ,
\ee
where $E_0$ is roughly the energy of the ground state. In the
case of the GUP in~(\ref{gup}), we can proceed as above and 
consider the term proportional to the square of the Planck 
length as a small correction to the standard formula. After 
trivial passages, one finds the order of magnitude of the 
correction to the ground state
\be
\Delta E_0 &\sim& 
L_{Pl}^2 \, m_e^3 \, e^8 \sim \nonumber\\
&\sim& m_e \, \left(\frac{E_0}{E_{Pl}}\right)^2
\sim 10^{-48} \; {\rm eV} \, ,
\ee
where $E_{Pl} = 1/L_{Pl} \sim 10^{19}$~GeV is the Planck energy.
On the other hand, if the GUP has the term linear in $L_{Pl}$, 
the induced energy shift of the ground state wuold be
\be
\Delta E_0 &\sim& 
m_e^{1/2} E_0^{1/2} \, \left(\frac{E_0}{E_{Pl}}\right)
\sim 10^{-23} \; {\rm eV} \, .
\ee
Such a correction is surely still too small to be experimentally
tested: at present, theoretical predictions and laboratory
measurements can agree at best at the level of some kHz,
about $10^{-11}$ eV, as in the case of the Lamb shift in the
hydrogen atom~\cite{codata}. However, the effect would be
anyway not so much suppressed as it is commonly thought and, 
hopefully, future investigations on the subject may point
to some phenomenon where corrections to the HUP could be
experimentally interesting in the case the leading order term 
is proportional to $L_{Pl}$, while could not in the
case the latter is proportional to $L_{Pl}^2$.

\section{Harmonic oscillator \label{sec-ho}}

Another interesting system with a lot of possible applications
is represented by the harmonic oscillator. For example, it can
provide a crude picture of heavy meson systems such as the
charmonium and the bottomonium~\cite{mesons}, where the 
confining force can be described by a linearly rising potential 
at large distances. The Hamiltonian is
\be
E = \frac{p^2}{2m_q} + \frac{1}{2} m_q \omega^2 r^2 \, ,
\ee
where $m_q$ is the constituent quark mass, i.e. 
$m_c \approx 1.3$~GeV for the $c$--quark and $m_b \approx 4.5$~GeV 
for the $b$--quark~\footnote{To be more precise, here $m_q$ wuold
be the reduced mass of the system, i.e. $m_c/2$ or $m_b/2$, but for
our order of magnitude estimate this is not relevant.}. From the 
HUP $p \gtrsim 1/r$, we find
\be
E \gtrsim \frac{1}{2 m_q r^2} + \frac{1}{2} m_q \omega^2 r^2 
\ge \omega \, .
\ee
$\omega$ is the binding energy of
the system and is roughly equal to the energy gap separating
adjacent levels, so $\omega \sim 0.3$~GeV. The correction of 
the standard GUP is
\be
\Delta E &\sim&
m \, \left(\frac{\omega}{E_{Pl}}\right)^2 \sim 
10^{-36} \; {\rm MeV} \, ,
\ee
while in the case the leading order correction to the HUP
is linear in $L_{Pl}$ we have 
\be
\Delta E &\sim&
m^{1/2} \omega^{1/2} \, 
\left(\frac{\omega}{E_{Pl}}\right)
\sim 10^{-17} \; {\rm MeV} \, .
\ee
Experimental uncertainties are typically something less than
1~MeV, but in the case of $J/\psi$ the accuracy is at the level
of $10^{-2}$~MeV~\cite{pdg}. Theoretical predictions are much 
more difficult than the hydrogen atom because are based 
on semi--phenomenological models. So, these quantum gravity 
effects are clearly well beyond experimental tests for the 
foreseeable future.

\section{Conclusions \label{sec-concl}} 

There are essentially no doubts that gravity implies a revision 
of the Heisenberg uncertainty principle, but it is always assumed
that the first order correction to the standard formula is 
proportional to the gravitational constant $G_N$, that is to the 
square of the Planck length $L_{Pl}$. In this paper I have 
shown that this is not necessarily true and that, once $L_{Pl}$
has acquired a clear physical meaning, associated with the 
minimum observable length of the theory, there are apparently no
reasons to exclude that the leading order correction is proportional 
to the first power of $L_{Pl}$. Such a result cannot be deduced
at a first approach, because common derivations of the Generalized
Uncertainty Principle use the standard formulation of quantum
mechanics and general relativity, where the Planck length is 
only the square root of the gravitational coupling constant $G_N$.
On the other hand, the emerging picture necessarily requires
a revision of the ingredients used to deduced it.
For example, the existence of a minimum length implies a
modification of the standard de Broglie relation
and this affects the final expression of the Generalized
Uncertainty Principle. Similar consequences may arise also from
some ``deformed'' energy--momentum conservation. 
Stringy effects at the Planck/string scale may also play
a fundamental rule. Unfortunately, the presence or absence 
of the linear term is model dependent and at present there are
no firm theoretical arguments that can suggest what really happens. 
On the other hand, such an observation can make the phenomenology 
more interesting and hopefully within the reach of future 
experiments. Lastly, it should be clear that, anyway, the Generalized
Uncertainty Principle in eq.~(\ref{gup}) cannot be an exact
formula, but at best a good approximation valid at some (not yet
well understood) regime.

\begin{acknowledgments}
I would like to thank Federico Urban 
for helpful comments and suggestions.
This work is supported in part by NSF under grant PHY-0547794 
and by DOE under contract DE-FG02-96ER41005.
\end{acknowledgments}

\end{document}